\documentstyle[preprint,pre,aps]{revtex}
\begin{document}
\draft

\title{Statistical  Mechanics of Shell Models for 2D-Turbulence}

\author{E. Aurell}
\address{Deptartment of Mathematics, Stockholm University
          S--106 91 Stockholm, Sweden}

\author{G. Boffetta}
\address{Istituto di Fisica Generale, Universit\`a di Torino,
         Via Pietro Giuria 1, I-10125 Torino, Italy}

\author{A. Crisanti}
\address{Dipartimento di Fisica, Universit\`a di Roma ``La Sapienza'',
         P.le Aldo Moro 2, I-00185 Roma, Italy}

\author{P. Frick}
\address{Institute of Continuous Media Mechanics,
         Russian Academy of Sciences, \\
         1, Acad.~Korolyov Str., 614061, Perm, Russia}

\author{G. Paladin}
\address{Dipartimento di Fisica, Universit\`a dell'Aquila
         Via Vetoio, Coppito I-67100 L'Aquila, Italy}

\author{A. Vulpiani}
\address{Dipartimento di Fisica, Universit\`a di Roma ``La Sapienza'',
         P.le Aldo Moro 2, I-00185 Roma, Italy}

\date{May 11, 1994}

\maketitle

\begin{abstract}
We study shell models that conserve the analogues of energy and enstrophy,
hence designed to mimic fluid turbulence in 2D. The main result is that
the observed state is well described as a formal statistical equilibrium,
closely analogous to the approach to two-dimensional ideal hydrodynamics of
Onsager, Hopf and Lee. In the presence of forcing and dissipation we observe
a forward flux of enstrophy and a backward flux of energy. These fluxes can
be understood as mean diffusive drifts from a source to two sinks in a system
which is close to local equilibrium with Lagrange multipliers (``shell
temperatures'') changing slowly with scale. The dimensional predictions on
the power spectra from a supposed forward cascade of enstrophy, and from one
branch of the formal statistical equilibrium, coincide in these shell models
at difference to the corresponding predictions for the Navier-Stokes and
Euler equations in 2D. This coincidence have previously led to the mistaken
conclusion that shell models exhibit a forward cascade of enstrophy.
We also study the dynamical properties of the models and the growth of
perturbations.
\end{abstract}


\section{Introduction}
\label{s:introduction}
The idea of shell models of
turbulence\cite{Obukhov71,DesnyanskiNovikov,%
Siggia77,Zimin,YamadaOhkitani87,JensenPaladin91,EggersGrossmann}
is to replace the fluctuations
of a turbulent field
in an octave of wave numbers $2^n<|k|<2^{n+1}$
by one or a few  representative variables. The range of
wave numbers is called
a shell, and the variables are called shell
variables.
The dynamics of the shell variables should be chosen so as
to preserve as many as possible of the qualitative features
of the full equations.
In this way, it is possible to describe  the
cascade processes of  inertially conserved
quantities  by a chaotic dynamical system with a
 limited number of degrees of freedom.
Predictions from dimensional analysis can then be tested with
far greater accuracy than in full simulations of the Navier-Stokes equations.

This approach has been particularly successful for
3D turbulence in the models
introduced by Gledzer\cite{Gledzer} and Yamada \& Ohkitani
\cite{YamadaOhkitani87} (see also \cite{JensenPaladin91}
or \cite{Pisarenko93}), which, following
the recent literature, we will refer to as the GOY
models.
Nevertheless it has recently become clear that the situation
is more complicated. Particular variants of the GOY
models have in fact stable fixed points, corresponding
to the Kolmogorov scaling law in 3D\cite{Biferale94}.
The dimensional predictions
for the energy spectra come out correctly in these models \cite{Biferale94},
but there are no intermittency corrections, and
the whole phenomenology of a turbulent state (e.g.
sensitivity to initial conditions, positive Lyapunov exponents)
is absent.

We will here report on another instance where shell models give
significantly different results than expected at
first: shell models designed
to model fluid turbulence in 2D.
We will show that these models are
perfectly well described by
a formal statistical mechanics, closely similar to the approach
of Onsager\cite{Onsager}, Hopf\cite{Hopf} and Lee\cite{Lee}
for inviscid 2D hydrodynamics
(see also the review \cite{KraichnanMontgomery}).
If energy and enstrophy are pumped into the system by an
external force and removed by viscous terms, we do observe
a net mean flux of both energy and enstrophy from the force
to the viscous sinks. These fluxes are shown to be analogous
to mean drifts proportional to the gradients of the
conjugate quantities, as in a system which is locally, but
not globally, in thermodynamic
equilibrium.
There is no need to invoke
a cascade, neither in the inverse direction (of energy)
nor in the forward direction (of enstrophy).
We show that in these shell models the dimensional predictions
on the energy spectrum from a supposed forward cascade of
enstrophy coincide with equipartition of shell enstrophies, which is
one of the branches of the statistical equilibrium.
In previous investigations \cite{YamadaOhkitani88a,Frick,FrickAurell}
the simpler explanation of the observed spectrum in terms of
a weakly perturbed equlibria was overlooked.

The paper is organised as follows: in section~\ref{s:models}
we review the standard phenomenology of 2D turbulence
and the minimal constraints which have to be put
on shell models to take into account the difference to 3D.
We present the shell models tested in the
simulations which are all extensions
and variations of the 2D GOY model studied by several
authors\cite{Gledzer,YamadaOhkitani88a,Frick,FrickAurell}.
In section~\ref{s:equilibrium}
we recall the statistical equilibrium approach, and apply it to determine
the scaling laws and the probability distribution functions of
the shell models.
In section~\ref{s:numerical} we present numerical results on
a steady forced state with enstrophy and energy output,
and compare them with the predictions
of section~\ref{s:equilibrium}.
In section~\ref{s:dynamical} we present the results on
dynamical properties, which
are quite different from the
results for 3D shell models, but consistent with what
 is expected from a system close to equilibrium.
In section 6 we summarize our results.

\section{2D turbulence and shell models}
\label{s:models}

The main particularity of 2D Euler equations is
that the vorticity is a material invariant,
i.e. conserved along stream lines.
This leads to
an infinite number of additional
integrals of motion, since the global integral of any
functional
of vorticity is inertially preserved.
In practice one habitually considers the lowest polynomial positive
invariant,
the enstrophy.
One of the main  reasons for this is  that the enstrophy is also an
invariant of the truncated Euler equations in
Fourier space\cite{KraichnanMontgomery}.
Although energy and enstrophy are both conserved globally,
their support in wave vector space
may change over time. The changes must be in opposite
directions, such that if
enstrophy
is transported towards small scales (large wave numbers)
energy is transported to large scales (small wave numbers).

The Navier-Stokes equations describe in addition to the
inertial motion
the action of external forcing and molecular viscosity.
The first major difference to the situation in 3D is
that in 2D molecular viscosity cannot remove energy efficiently,
because enstrophy contains two more derivatives, is dissipated
first, and energy is then constrained to flow to large scales.
This implies that in numerical investigations
 an infra-red viscous term $\nu'(-\Delta)^{-\gamma}$
is necessary to remove energy at large scales:
\begin{eqnarray}
\partial_t \vec u + (\vec u\cdot\vec\nabla)\vec u &=& - \vec\nabla P
+ \nu\Delta\vec u - \nu'(-\Delta )^{-\gamma}\vec u + \vec F \\ \nonumber
\vec\nabla\cdot\vec u = 0\quad &\, &\quad\vec\nabla\cdot\vec F = 0
\label{NavierStokes2D}
\end{eqnarray}
The exponent $\gamma$ is rather arbitrary.
 For instance, $\gamma$ in (1)               
equal to zero corresponds to linear friction, which
describes the physical situation of a
fluid flow in thin films with viscous drag in boundary
layers.
If one is mainly interested in the backward energy cascade
ranges it is better to take a sharper artificial infra-red
viscosity, for which the action is more concentrated
in a small range of wave number.
This procedure is standard in numerical investigations
of the 2D Navier-Stokes equations
(see e.g. \cite{FrischSulem,Borue,KraichnanMontgomery}).

We will from now on assume that the force acts only on scales
with typical wave number $k_{f}$, and that the molecular and
infra-red viscosities act on scales respectively much smaller
and much larger. The spectrum can then be divided into five
ranges:
the energy dissipation range, where the infra-red viscosity
acts effectively; the inverse inertial
transport range, where energy
flows towards large scales; the scale of the forcing;
the forward inertial transport range, where enstrophy flows towards
small scales; and the enstrophy dissipation range, where
molecular viscosity acts effectively.

The first attempts to predict the behaviour of hydrodynamics
in 2D used a formal statistical mechanics analogy for the
Euler equations. We review this approach in
section~\ref{s:equilibrium}. A later approach due to Kraichnan and Batchelor
 proceeds along the lines of the Kolmogorov cascade picture
for 3D\cite{Kolmogorov41}.
For the forward transport range, a cascade of enstrophy implies
by dimensional arguments that the energy spectrum should
be $E(k)\sim k^{-3}$. Similarly, for the inverse transport
range, an inverse cascade of energy gives
$E(k)\sim k^{-{5\over 3}}$\cite{Batchelor,Kraichnan,KraichnanMontgomery}.

In 3D there is strong
experimental support of the cascade picture\cite{MoninYaglom},
albeit with
intermittency corrections\cite{Anselmet84,Castaing90}.
The evidence for cascade processes in 2D is not as clear.
In the forward range,
 a energy spectrum steeper than $k^{-3}$ implies that the most
important interactions are not local in wave number space.
A scaling $k^{-3}$ is marginal, and the
dimensional predictions based on a
cascade process with local interactions in $k$-space
are questionable. In fact, most
 numerical investigations  of the 2D Navier-Stokes equations
 report spectra which are significantly steeper than
 $k^{-3}$\cite{BasdevantCouder,Benzi84}.
However, all numerical investigations of the full Navier-Stokes
equations suffer from a limited
range of wave numbers for the inertial ranges.
The $k^{-3}$ energy spectrum can therefore
not conclusively
be ruled out, while the prediction of
the statistical equilibrium approach, which is $k^{-1}$,
is clearly excluded by the numerical observations.

The true state of the 2D Navier-Stokes equations
in the inverse cascade range
is unclear.
Most simulation have reported a $k^{-{5\over 3}}$ spectrum in
agreement with the dimensional
 predictions \cite{FrischSulem,SmithYakhot}.
The latest investigation over much longer time-scales
instead reports a $k^{-3}$ spectrum extending to
scales about one order of magnitude
larger than the forcing, and then a spectrum clearly
much flatter than $k^{-{5\over 3}}$\cite{Borue}.
In the inverse range the prediction from
statistical equilibrium, which is $E(k)\sim k$,
does therefore not seem to be completely ruled out.

The shell models we consider should be taken as the
simplest systems that
share the property of having two inertially conserved
quadratic quantities.
They are systems of coupled
ordinary equations which may in general be written as:
\begin{equation}
{d\over {dt}} W_n = \sum\limits_{m,l} X_{nml} W_m W_l - K_n W_n + F_n
\label{shellZ}
\end{equation}
where $X$ indicates the nonlinear interactions, $K$ the viscous
forces and $F$ an external force.
In (\ref{shellZ}) we have suppressed an index
of shell variables within one shell. In the extreme case
considered in the literature \cite{EggersGrossmann} this can be
a moderately large number, of the order of $10^2$.
In the cases we consider, we will have one or a few complex shell
variables per shell, with nonlinear terms that are compactly
expressed using the complex conjugates $W_n^*$.

We use the convention that the shell variable $W_n$ is like
a shell vorticity, so (\ref{shellZ}) is a model of the
vorticity equation. The quantities that should be
inertially conserved are then:
\begin{eqnarray}
\hbox{``enstrophy'';}\qquad\qquad \Omega &=& \sum_n |W_n|^2 \\
\hbox{and ``energy'';}\qquad\qquad E &=& \sum_n |{1\over{k_n}}W_n|^2 ,
\end{eqnarray}
where $k_n$ is the typical wave number of shell $n$.
We will only consider the case where the $k_n$'s are
spaced in octaves, i.e. $k_n=k_o\,2^n$.

The dimensional predictions are based on the assumption
 of cascade processes and from the observation that the fluxes of
energy $\Pi_n(E)$ and of
enstrophy $\Pi_n(\Omega)$ through the $n^{\rm th}$ shell
are given by certain third order correlation
functions\cite{Pisarenko93}. For
the 2D GOY model (see equation (\ref{Yam-Ohk2D})) they read in detail
\begin{eqnarray}
\label{GOYenstrophyflux}
\Pi_n(\Omega) &=& \langle\hbox{Re}(4W_n W_{n-1} W_{n-2} -
W_{n+1}W_nW_{n-1})\rangle\\
\label{GOYenergyflux}
\Pi_n(E) &=& {4\over{k_n^2}} \langle\hbox{Re}
(W_n W_{n-1} W_{n-2} - W_{n+1}W_nW_{n-1})\rangle
\end{eqnarray}
If there is a net forward transport of enstrophy
from the force to the enstrophy dissipation,
the correlation functions (\ref{GOYenstrophyflux})
(and the analogous expressions for the other models)
must be constant in the forward transport range.
That gives the dimensional estimate
\begin{equation}
\langle|W_n|\rangle~\sim~\epsilon_{\omega}^{2\over 3}\, k_n^0,
\label{enstrophycascadeestimate}
\end{equation}
where
$\epsilon_{\omega}$ is the mean dissipation of enstrophy
per unit time.
Similarly in the inverse transport range one gets the
estimate
\begin{equation}
\langle|W_n|^2\rangle~\sim~\epsilon^{2\over 3}\, k_n^{4\over 3},
\label{energycascadeestimate}
\end{equation}
where $\epsilon$ is the mean dissipation of energy per unit time.
The shell energy estimates are respectively
$E_n\sim k_n^{-2}$ and $E_n\sim k_n^{-{2\over 3}}$,
which is the same as the cascade estimates
$E(k)\sim k^{-3}$ and $E(k)\sim k^{-{5\over 3}}$ for the
full Navier-Stokes equations integrated
over one octave in wave number space.

\subsection{The GOY models in 2D}
\label{ss:yamada-ohkitani}
The GOY models are usually written in terms
of complex shell velocities as
\begin{equation}
{d\over {dt}}U_n = i k_n (a_n U_{n+1}^* U_{n+2}^*
+ {{b_n}\over 2} U_{n-1}^* U_{n+1}^* + {{c_n}\over 4} U_{n-1}^* U_{n-2}^*) -
\nu k_n^2 U_n + f_n
\label{Yam-Ohk}
\end{equation}
We can transform (\ref{Yam-Ohk}) to (\ref{shellZ}) by absorbing
a factor $k_n$ in the shell variables.
The particular form of the right-hand side of (\ref{Yam-Ohk})
ensures that the phase-space volume
\begin{equation}
dV = dU_1\wedge dU_1^* \wedge\ldots\wedge dU_n\wedge dU_n^*
\end{equation}
is inertially conserved.
The parameters values for which (\ref{Yam-Ohk}) conserve energy
are given up to an overall factor by
\begin{equation}
a_n =1\quad b_n = -\epsilon\quad c_n = -(1-\epsilon)
\label{Yam-Okh}
\end{equation}
The bifurcations and transitions to chaos in (\ref{Yam-Ohk})
when changing $\epsilon$ have recently been investigated in
\cite{Biferale94}. If $\epsilon$ is greater than 1 there is a second
conserved quantity; $\sum k_n^{2\alpha}|U_n|^2$, the exponent $\alpha$
changing with $\epsilon$\cite{OhkitaniYamada}.

Enstrophy conservation, i.e. $\alpha=1$, is realized at the particular value
$\epsilon$ equal to $1.25$. The 2D GOY
equations written for shell vorticity variables hence read
\begin{equation}
{d\over {dt}} W_n =  (W_{n+1}^* W_{n+2}^* -5W_{n-1}^* W_{n+1}^*
+ 4W_{n-1}^* W_{n-2}^*) -
\nu k_n^2 W_n -\nu' k_n^{-2\gamma}W_n + F_n
\label{Yam-Ohk2D}
\end{equation}
For convenience we have in (\ref{Yam-Ohk2D}) rescaled the $W_n$'s
to absorb an overall factor ${i\over 8}$, and added an artificial
infra-red viscosity.

In \cite{Frick} a class of models were proposed that contain
GOY-like interactions between triads of shells further
apart. These models were further studied in the range
of forward transport in \cite{FrickAurell}.
For interactions decaying sufficiently fast with distance
between interacting shells they behave quite similarly
to the 2D GOY model.

\subsection{The coupled GOY model}
\label{ss:vap}
We motivate this model by the observation that the GOY models
make no difference between velocity and vorticity, which only
differ by a scale factor. It could therefore be hoped that a shell
model which preserves some trace of the vector structure of the
velocity field will be
in a qualitative sense closer to the Navier-Stokes
equations.

In 2D we can write the
vorticity
equation in Fourier space as
\begin{eqnarray}
{d\over {dt}}\, \hat\omega(\vec k) =  \sum_{\vec k + \vec k' + \vec k'' = 0}
{{ \vec k'\times \vec k''}\over{|\vec k'|^2}}
\,\hat\omega(\vec k')\, \hat\omega(\vec k'')
- \nu |\vec k|^2 \hat\omega(k) +F(k)
\label{vort2D}
\end{eqnarray}
An obvious consequence of (\ref{vort2D}) is that two parallel
vectors do not drive a third.
We can model (\ref{vort2D}) by taking
few shell variables, $W_{n,j}$, $j=0,1,\dots $,
per shell, associate to each
a wave vector $\vec k_{n,j}$ with
length $k_n$ and direction $\hat e_j$,
and look for equations in the
form
\begin{equation}
\begin{array}{l}
{\displaystyle d\over \displaystyle  dt} W_{n,j} =
\sum\limits_{(n',j'),(n'',j'')}
C_{(n,j),(n',j'),(n'',j'')}\,
{\displaystyle \vec k_{n',j'}\times \vec k_{n'',j''} \over
 \displaystyle |\vec k_{n',j'}|^2}\,
W_{n',j'}^* W_{n'',j''}^* \\ \\ \qquad\qquad
-\nu k_n^2 W_{n,j} -\nu' k_n^{-2\gamma}W_{n,j} + F_{n,j}
\end{array}
\label{complicatedshell}
\end{equation}
with some interaction coefficients $C$ chosen to preserve
energy and enstrophy.

A simple implementation of (\ref{complicatedshell}) is to take
wave vectors in a hexagonal pattern,
$\hat e_j~=~(\cos(2\pi j/3),\sin(2\pi j/3))$ for $j=0,1,2$.
One possible set of interactions are then the same as in
the 2D GOY
model, but going only between triples
of shell variables having different
directions:
\begin{equation}
\begin{array}{l}
{\displaystyle d\over \displaystyle dt} W_{n,j} =
                   (W_{n+1,j+1}^* W_{n+2,j+2}^*
		-5 W_{n+1,j+1}^* W_{n-1,j+2}^*
		+4 W_{n-2,j+1}^* W_{n-1,j+2}^* )\, +
 \\  \qquad\qquad
		(W_{n+2,j+1}^* W_{n+1,j+2}^*
		-5 W_{n-1,j+1}^* W_{n+1,j+2}^*
		+4 W_{n-1,j+1}^* W_{n-2,j+2}^* )\,
 \\  \qquad\qquad
-\nu k_n^2 W_{n,j} -\nu' k_n^{-2\gamma}W_{n,j} + F_{n,j}
\end{array}
\label{coupledYam-Ohk}
\end{equation}
In (\ref{coupledYam-Ohk}) we have used the notation that
$W_{n,j+3}= W_{n,j}$.
We will call (\ref{coupledYam-Ohk}) the coupled GOY model.
It has some
moderate numerical advantages over the original
GOY model.
Due to more variables
per shell and more couplings, the coupled GOY model
equilibrates faster.
In the original GOY models with a steady
force one observes that the shell energy spectrum in the
inertial range exhibits oscillations superimposed on a mean
power-law\cite{Pisarenko93}.
We find no trace of such oscillations in (\ref{coupledYam-Ohk}),
and conclude that three
variables per shell is enough to sufficiently
randomize the system
and remove these undesired oscillations.

We have used forcing terms acting on shell $0$ of the forms:
\begin{eqnarray}
\label{fluxconst}
F_{0,j} &=& {{\eta}\over{\sum\limits_{j=0,1,2} |W_{0,j}|^2}} W_{0,j} \\
\label{almostconst}
F_{0,j} &=& {{\eta}\over{\sum\limits_{j=0,1,2} |W_{0,j}|^2}} W_{0,j}^*
\end{eqnarray}
The advantage of (\ref{fluxconst}) and (\ref{almostconst})
is that the fluxes of energy
and enstrophy from the force are fixed to be proportional
(in the case (\ref{fluxconst}) equal) to $\eta$.
For most of our simulations
we have used artificial viscous terms like
$-\nu' k_n^{-2} W_n$, i.e. $\gamma = 1$.

\section{The statistical mechanics approach}
\label{s:equilibrium}
We here review the approach from equilibrium statistical mechanics
of Onsager\cite{Onsager}, Hopf\cite{Hopf} and Lee\cite{Lee}. For a later review
see \cite{KraichnanMontgomery}.

We consider the equations of two-dimensional hydrodynamics
in Fourier space, with cut-offs $k_{\rm min}$ and
$k_{\rm max}$.
The lower cut-off can be realised physically by enclosing the system
in a finite container. We need
also the upper cut-off to avoid
the ultra-violet catastrophe of classical continuous fields.
As the system leaves inertially invariant
the two quadratic forms energy and enstrophy, the canonical
ensemble
distribution function for the Fourier components of vorticity
is
\begin{equation}
P(\hat\omega(\vec k)) \propto \exp \bigr[-(\beta_1\,|\hat\omega(\vec k)|^2
       + \beta_2\, |\vec k|^{-2}|\hat\omega(\vec k)|^2 ) / 2 \bigl]
\label{canonicaldistribution}
\end{equation}
In (\ref{canonicaldistribution}) we have two Lagrange multipliers,
$\beta_1$ and $\beta_2$, or, alternatively,
an ``enstrophy temperature'' $T_1$, equal to $1 / \beta_1$,
and an ``energy temperature'' $T_2$, equal to $1 / \beta_2$.
The second order moment of (\ref{canonicaldistribution})
is
\begin{equation}
\langle |\hat\omega(\vec k)|^2\rangle = {{T_1}
\over{ 1+ {{T_1}\over{T_2}} |\vec k|^{-2} } }
\label{secondmoment}
\end{equation}
which, depending on $|\vec k|^2$ and the temperatures, separates into two
branches
\begin{equation}
\langle |\hat\omega(\vec k)|^2\rangle\, \sim\,
\left\{
	\begin{array}{ll}
	T_2\, |\vec k|^2 \qquad &\hbox{if $|\vec k|^2 \ll {T_1\over T_2}$} \\
	T_1  \qquad        &\hbox{if $|\vec k|^2\gg{T_1\over T_2}$ }
	\end{array}
\right.
\label{secondmomentasympt}
\end{equation}
The energy spectra in these two ranges are
$E(k)\sim k$ and $E(k)\sim k^{-1}$, respectively.
As discussed above at least the second prediction
is clearly in disagreement with all
numerical investigations of the 2D Navier-Stokes
equations\cite{FrischSulem,SmithYakhot,Borue,BasdevantCouder}.

The argument for (\ref{canonicaldistribution})
translates immediately to the shell models considered
in section~\ref{s:models}. We then have the prediction
that the shell variables are uncorrelated Gaussians with
widths
\begin{equation}
\langle |W_n|^2\rangle\, \sim\,
\left\{
	\begin{array}{ll}
	T_2\, k_n^2 \qquad &\hbox{if $k_n^2 \ll {{T_1}\over{T_2}}$ } \\
	T_1  \qquad        &\hbox{if $k_n^2 \gg {{T_1}\over{T_2}}$ }
	\end{array}
\right.
\label{shellprediction}
\end{equation}
The two branches in (\ref{shellprediction})
simply correspond to
shell energy equipartition and shell enstrophy equipartition.
The shell energy prediction in the shell enstrophy
equipartition branch is $E_n\sim k_n^{-2}$.
We thus have that in 2D shell models, but not in 2D ideal
hydrodynamics, the energy spectrum of
one branch of a formal statistical equilibrium and of
 an assumed forward cascade of enstrophy exhibits the same scaling law.

We continue with some elementary considerations
from non-equilibrium
statistical mechanics\cite{LandauLifshitz10}.
Suppose that a physical system is not in thermal equlibrium, but the
distribution functions at any point are close to the equilibrium
distribution, with temperatures that change slowly in
space. A heat flux is then set up, which acts
to restore global equilibrium. In other words, the induced
flux is
\begin{equation}
\vec \Pi_Q\, = \, -\kappa \nabla T
\end{equation}
where the proportionality
constant $\kappa$ is the heat conduction coefficient, generally
dependent on temperature.

We now translate this picture to the shell models.
For definitiveness we consider
the range of forward transport of enstrophy.
The local ``shell temperatures'' of this non-equilibrium
system are not defined {\it per se},
but may be specified by
the expectation values
\begin{equation}
T_1(n)\quad = \quad \langle|W_n|^2\rangle
\end{equation}
where the average is taken with respect to the
the stationary distribution of the $W_n$'s.
In first approximation, it is assumed that this joint distribution
 factorizes into uncorrelated
Gaussians with widths $T_1(n)$.
Then, if the shell temperatures of two nearest neighbour shells
differ, we expect a flux of enstrophy through the $n^{\rm th}$ shell.
\begin{equation}
\Pi_n (\Omega)\, =\, - \kappa_1 (T_1(n+1)-T_1(n))
\end{equation}
where we can call the transport coefficient
$\kappa_1$ the ``shell enstrophy
conductivity constant''.
In a steady state of forward flux
of enstrophy, one thus has
\begin{equation}
T_1(n+1)=T_1(n) - {{\epsilon_{\omega}}\over{\kappa_1}}
\label{transportcoeff}
\end{equation}
where $\epsilon_{\omega}$ is the mean dissipation of enstrophy per
unit time.

The temperature dependence of $\kappa_1$ can be estimated
as follows. The nonlinear inertial term in the shell model
equations  plays the role of the collision operator in
kinetic theory. The transport coefficient is determined by
the linearized collision operator. In the shell models
the nonlinear terms are quadratic, and the linearization
hence linear in the shell amplitudes. We therefore expect
\begin{equation}
\kappa_1 \sim \sigma \sqrt{T_1}
\label{gaskinetics}
\end{equation}
with some proportionality constant $\sigma$, as
in the kinetic theory of gases.

When all the temperature increments are supposed small we have
the prediction
\begin{equation}
T_1(n) \sim  [(T_1(n_f))^{3\over 2} -
{{3\epsilon_{\omega}}\over{2\sigma}}(n-n_f)]^{2\over 3}
\label{preasymptotics}
\end{equation}
where $n_f$ is the shell of the forcing.
At the dissipative end $T_1(n)$ eventually
becomes much smaller than unity. We may therefore rewrite
(\ref{preasymptotics}) as
\begin{equation}
T_1(n) \sim  \left[{{3\epsilon_{\omega}}\over{2\sigma}}
(n_{\rm diss}-n)\right]^{2\over 3}\qquad\qquad
\nu k_{n_{\rm diss}}^2 \sim 1
\label{preasymptotics2}
\end{equation}
We see that by coincidence the dependance on
$\epsilon_{\omega}$ is also as predicted by the cascade
picture. The spectrum however depends logarithmically on
$\nu$ through $n_{\rm diss}$. This contradicts
the assumption of a cascade process independent of viscosity.

Finally, over small increments the transport coefficient
are slowly varying, and
we have the shell model equivalent of Fourier's law of
heat conduction:
\begin{equation}
T_1(n) \sim  T_1(n_0) - {{\epsilon_{\omega}}\over{\kappa_1}}(n-n_0)
\qquad\qquad (n-n_0) \ll (n_{\rm diss} - n_f)
\label{fourierslaw}
\end{equation}

\section{Numerical results on the steady state}
\label{s:numerical}
In this section we present numerical results on the coupled
GOY model (\ref{coupledYam-Ohk}).
The first results, figures \ref{f:structfunc2} to \ref{f:energy_flux}
are from one run of length
$3\cdot 10^5$ in time. Statistics was taken with frequency
1 Hz, about once per shell turnover time in the range
of forward transport.
By convention the force acted on shell
zero. The viscosity was $\nu=10^{-24}$. Practically all
dissipation of enstrophy occurred in shells $34$ to $42$.
The artificial viscous term was of the type $-\nu' k_n^{-2} W_n$
with $\nu'=10^{-15}$. The energy output occurred in shells
$-21$ to $-14$.
We used the forcing of type (\ref{almostconst}) with
$\eta$ equal to one. The observed mean fluxes of enstrophy
and energy were approximately $0.075$.
The integration method was the slaved leap-frog scheme of
\cite{Pisarenko93} with time-step $0.003$.

Let us first look at figure~\ref{f:structfunc2}
which shows the second moment of shell vorticity vs.
shell number.
The dominant overall feature is one branch at positive $n$'s
which is nearly flat, i.e. implies $\langle |W_n|^2\rangle\sim k_n^0$,
and one branch at negative $n$'s which is closely fitted
by $\langle |W_n|^2\rangle\sim k_n^2$. Both results are in agreement
with the predictions of a statistical equilibrium as in
section~\ref{s:equilibrium}.
Let us add here that according to figure~\ref{f:structfunc2}
the characteristic times
in the range of forward transport of enstrophy
are about constant. This run
therefore represents about $10^5$ turnover times in all
the forward range, and there are indeed only small fluctuations.
In contrast, in the inverse range the characteristic times
are proportional to $k_n^{-1}$, which means that we have
about one turnover time
in the energy dissipation range.
Looking closer on figure~\ref{f:structfunc2} we clearly
see that the spectrum is in the range of forward transport
is not quite flat, and that the ``$k_n^0$-range'' seems to extend
about five shells to the left of the force.
The second phenomenon is difficult to explain in the framework of
 the cascade picture. It is not a problem in the statistical
equilibrium picture if we assume that the shell temperatures
of enstrophy and energy are such that the bend in the curve
occurs somewhat to the left of the force.
The first phenomenon was observed
in \cite{FrickAurell},
and there attributed to a non power-law correction on top of
a power-law with an exponent close to zero.
Systematic investigations
on the Yamada-Ohkitani models\cite{Boffettaunpub}
gave only a pre-asymptotic
correction, and thus an asymptotic power-law
$k_n^0$.
Figure~\ref{f:stateq} shows the second moments of the shell
variables in real scale, instead of logarithmic scale as
in figure~\ref{f:structfunc2}. A dominant linear trend is
quite clear. The deviations from a pure power-law
in figure~\ref{f:structfunc2} are hence as predicted by
(\ref{fourierslaw})
from the picture of weakly perturbed equilibria.



Figure~\ref{f:gaussian-forward} shows the numerically
observed distributions in the forward range, compared
with a Gaussian. The fit is clearly good.
Figure~\ref{f:kurtosis} compares the numerically observed
kurtosis of the shell enstrophies
to the estimate that the three variables in one shell are
uncorrelated Gaussians with the same variance.
Again the fit is quite good.
This analysis in the inverse range leads to similar scenario.
Taken together these results imply that to good
approximation there are no observable
correlations between the shell
variables within one shell. We have also checked directly that
this is true. In addition there are no measurable correlations
between shell variables in two nearest neighbour  shells.

Figure~\ref{f:enstrophy_flux} shows the mean enstrophy flux
and the mean squared enstrophy flux.
The curve of mean enstrophy flux should be flat, and is so to
a much better approximation than the second moment
in figure~\ref{f:structfunc2}.

Figure~\ref{f:enstrophy_flux} shows
that the standard deviation of the flux is about $50$ times higher
than the mean value in the beginning of the range of forward
transport, and then decreases, but stays above the mean flux
all up to the dissipative range.
The mean flux shows a clear downwards
jump from shell one to shell zero,
and then a fast decrease for negative $n$'s.
In this range the eventual
mean transport can only vanish, and the non-zero values represent
fluctuating incremented averages taken over finite time.


Figure~\ref{f:energy_flux} shows the mean energy flux
and the mean squared energy flux.
Towards the energy dissipation range the mean flux is to
good approximation constant, and agrees with the mean energy
infused to the system per unit of time.
At the top value around five shells to the
left of the forcing, the standard
deviation of the energy flux is about $10,000$ times
larger
than the mean value.
{}From the value of the variance of the flux, about $10^6$,
and the number of measurements, $3\times 10^5$,
one sees that rise of the curve of the energy flux below the
top value of the flux squared curve
is a statistical measurement
error.
We have checked that this explanation is
correct by comparing with other runs with higher artificial
viscosity for which the range of inverse energy
transport is
shorter, and the times to reach an asymptotic state at the
energy dissipation scale is shorter.

The second result are direct tests of equation~(\ref{gaskinetics}),
from different runs with different viscosities and different
forward enstrophy flux. Figure~\ref{f:gaskinetics} shows
that $\kappa_1$ scales to good approximation as $\sqrt{T_1}$
as expected by our theoretical arguments.

\section{Dynamical behaviour}
\label{s:dynamical}

In this section we briefly discuss the "instability properties" of the
2D shell models (\ref{coupledYam-Ohk}) by studying the
behavior of the Lyapunov
exponent $\lambda$ as function of the Reynolds number $Re$, the fluctuations
of the effective Lyapunov exponent $\gamma_{\tau}$, the tangent vector
and the spatial spreading of a small perturbation initially localized on
a given shell.

Given a dynamical system described by a set of differential equations
\begin{equation}
  \frac{dx}{dt} = F[x(t)]
\nonumber
\end{equation}
the response of the system to a perturbation $\delta x(t+\tau)$ of its state
at time $t$ after a delay $\tau$ is measured by the error growth rate
\begin{equation}
  R_{t}(\tau) \equiv {|z(t+\tau)|\over|z(t)|}
              \simeq {|\delta x(t+\tau)|\over|\delta x(t)|}
\nonumber
\end{equation}
where $z$ is the tangent vector obeying the evolution equation
\begin{equation}
 \frac{dz_i}{dt} = \sum_{j}\, {\partial F_i[x(t)]\over\partial x_j}\,z_j.
\end{equation}
By definition the largest Lyapunov exponent is
\begin{equation}
 \lambda = \lim_{\tau\to\infty}\, {1\over \tau}\,
             \langle\ln R_{t}(\tau)\rangle
\end{equation}
where the angular brackets denote a time average along the trajectory.
The Oseledec theorem \cite{Oseledec} ensures that if the average is
removed, then for
almost all initial conditions one obtains the same value for
$\lambda$.

The exponent $\lambda$ gives a global characterization of the
``instability'' of the trajectory. Local informations can be
obtained from the effective Lyapunov exponent $\gamma_{\tau}(t)$
defined as
\begin{equation}
 \gamma_{\tau}(t) = {1\over\tau}\,\ln R_{t}(\tau)
\end{equation}
and from its fluctuations\cite{Benzi85}.

The scenario which emerges is completely different from that observed for
the 3D GOY model studied in
\cite{JensenPaladin91}. The main results for the 3D
model are that:
\begin{center}
\begin{enumerate}
\item the Lyapunov exponent $\lambda$ increases with the Reynolds number:
\begin{equation}
\lambda \sim Re^{\alpha}\ \hbox{\rm with}\ \alpha\simeq 0.46
\end{equation}
in good agreement with the prediction of the multifractal
 generalization\cite{crisanti93}
of a Ruelle argument\cite{Ruelle}.

\item the effective Lyapunov exponent $\gamma_{\tau}$ exhibits strong
fluctuations, at increasing $Re$. Its variance $\mu$ scales like
\begin{equation}
\mu=\lim_{\tau \to \infty}
   \tau \langle(\gamma_{\tau} - \lambda)^2\rangle
 \sim Re^{\beta} \ \hbox{\rm with}\ \beta\simeq 0.8
\end{equation}
so that the ratio $\mu/\lambda$, which gives a quantitative
measure of the intermittency level\cite{Benzi85}, diverges
 with the Reynolds number.

\item
the tangent vector during the intermittent bursts (of energy and chaoticity)
 is strongly localized
on the shells corresponding to the Kolmogorov scale, while in the
laminar phase spreads over all the inertial range.
In addition there is a strong correlation between the intermittent bursts
of energy dissipation, large fluctuations of the effective Lyapunov
exponent and the localization of the tangent vector on the Kolmogorov
length.

\item
there is a backward cascade for the propagation of a perturbation from
small scales to large scales in qualitative agreement with the
phenomenological scenario proposed by Lorenz \cite{Lorenz69}
\end{enumerate}
\end{center}

The 2D shell models present a completely different behaviour. The numerical
study of the models introduced in this paper
reveals the following
scenario:

\begin{enumerate}
\item
the Lyapunov exponent depends very weakly on the Reynolds number:
\begin{equation}
\lambda \sim \ln Re
\end{equation}
see figure~\ref{f:Lyapunov}. It is worth noting that even in the framework
 of the Kraichnan-Batchelor theory, one has the same prediction
 for 2D Navier-Stokes equations,
by assuming that the Lyapunov exponent is proportional to the
smallest characteristic time of the system.

\item
the effective Lyapunov exponent has small fluctuations and
$\mu\ll\lambda$.

\item
the tangent vector is
concentrated at the scale of the forcing.

\item
a small perturbation initially concentrated
about a given shell does not propagate by an inverse cascade mechanism,
but diffuses through the shells.
In Fig.~\ref{f:spreading} we report the spreding of a perturbation initially
located on a small number of shells. The quantity shown in figure is defined as
\begin{equation}
\delta W_n = \left[
                \sum_{j=0,1,2} (W_{n,j} - W_{n,j}')^2
             \right]^{1/2}
\label{eq:spread}
\end{equation}
where at the initial time $W_{n,j}'$ differs from $W_{n,j}$ of a quantity
$\delta$ on shells $n = n_1,\dots,n_2$.
In the case reported in Fig.~\ref{f:spreading} we used $n_1=-6$, $n_2=-5$
and $\delta=10^{-10}$. However the curves are rather insensitive to the
initial location of the perturbation.

\end{enumerate}



The above results provide a clear evidence
 the ``instability'' behavior
in the 2D shell models is very different from those of 3D shell models.
Let us remark that  the dynamical  features observed
for equation~(\ref{coupledYam-Ohk})
are rather
close  a scenario expected in equlibrium statistical mechanics.

\section{Discussion}
\label{s:discussion}
We have presented numerical results on a wide class of shell
models of 2D turbulence which can be simply
and coherently explained by a formal non-equilibrium statistical
mechanics, close to local equilibrium.
 The presence of an inverse cascade of energy
can be ruled out already by analyzing
 the second moments of the shell variables.
In the forward inertial  range
 a direct cascade of shell enstrophy would predict
the correct scaling behaviour of the
second moments with shell wave number and mean dissipation
of enstrophy. However, we also find that the spectrum
rises as the viscosity decreases. This rules
out a cascade process which assumes a state independent
of viscosity in the low viscosity limit.

The weakly perturbed equilibria predicts Gaussian probability
distribution functions of the shell variables, and the pre-asymptotic
corrections to the power-law in the forward range.
Let us remark that these two observations are difficult to
reconcile in a cascade picture where Gaussian probability distribution
functions are possible if there
is only one scaling exponent, ``unifractality'', \cite{Benzi91}.
However, in absence of multifractality, it is hard to imagine
 mechanisms which are able to produce
corrections to scaling  such as
 pseudo-algebraic power laws, the so-called multiscaling
\cite{FrischVergassola}.

Finally, the mean values of the fluxes of both energy and enstrophy
are always much smaller than the standard deviations.
The fluxes are thus always small corrections superimposed on
an mean randomly fluctuating state.

For the full hydrodynamic equations in 2D
a state of equilibrium
is ruled out in the range of forward transport. We therefore
conclude that the models studied in this paper have very little
to do with 2D turbulence.

A natural question at this point is: why have shell models
for 3D turbulence given reasonable results, while the 2D
models have not? A qualitative answer goes as follows: the
statistical equilibria picture should be relevant if the time-scales
of relaxation to local equilibrium are faster than the
timescales of transport of the conserved quantities to the viscous
sinks.
The time-scale for relaxation to local equilibrium can
be estimated as the local shell turnover times, which
in 3D shell models decrease with
shell number (as $t_n~\sim~k_n^{-2\over 3}$ at the Kolmogorov fixed
point). The time scale for transport to the energy dissipation
range from a shell in the forward transport
range can then be estimated as the geometric sum of the local
turn-over times up to the energy dissipation shell.
Hence both time scales are of the same order, and it is unlikely
that 3D shell models display statistical equilibrium. And indeed
a cascade scenario is observed.

On the contrary, in 2D shell models the local turnover times
in the forward range are constant, and the time to transport enstrophy
to the dissipation range is proportional to how far away in shells
that is. Therefore local statistical equilibrium has a chance to
develop.

The question then arises why this argument does not
work for the 2D Navier-Stokes equations, where, in the Batchelor-Kraichnan
cascade picture, the timescales in the forward range are also
constant. One possible explanation is that one has to take into
account the full nonlinearities of the
Navier-Stokes equations with the non-local transfer of energy and
enstrophy.
A more interesting possibility is that the reason is that there
are more states at high wave-numbers, and that the system gains
entropy locally by transporting in that direction.
It would then make sense to study simplified models but where
the number of states increases with shell number. Numerical
experiments on one such model \cite{Aurell} lend support
to this idea.
In the inverse transport range the number of states decrease
towards the energy dissipation range. A speculative conclusion
is of this work is then that
perhaps the true state of the 2D Navier-Stokes equation
is a forward cascade of enstrophy, or a non-local transfer
with spectrum steeper than $k^{-3}$, on scales smaller than
the forcing, but a
formal statistical equilibrium
on scales larger than the forcing.
The latest numerical simulations by Borue\cite{Borue}
seem to agree with this picture.

\acknowledgments
This work was supported by
the Swedish Natural Science Research Council under contract
S--FO 1778-302~(E.A.), by the French Government Program
``Relance de l'Est'' (P.F.), and by INFN (Iniziativa Specifica Meccanica
Statistica FI3).
E. A. thanks the Dipartimento di Fisica,
Universit\`a di Roma ``La Sapienza'' for hospitality and
financial support.
G. B. thanks the ``Istituto di Cosmogeofisica del CNR'', Torino, for
hospitality.
P.F. thanks the Laboratoire de M\'et\'eorologie Dynamique,
Ecole Normale Superieure, Paris for hospitality.

\begin{figure}
\caption{The second order structure function
         $\omega_2(n)= \left\langle\sum_{j} |W_{n,j}|^2\right\rangle$
         of shell
         vorticity as a function of shell number, logarithmic scale.
        }
\label{f:structfunc2}
\end{figure}

\begin{figure}
\caption{The second order structure function of vorticity
         as a function of shell number in the range of forward
         transport of enstrophy, real scale.
        }
\label{f:stateq}
\end{figure}

\begin{figure}
\caption{Probability distribution function of the real parts
         of the shells variables $W_{5,0}$ and $W_{30,0}$
         in the range of forward transport of enstrophy. The full
         line is a Gaussian distribution with the same variance.
        }
\label{f:gaussian-forward}
\end{figure}

\begin{figure}
\caption{The kurtosis $\omega_4(n)/\omega_2(n)^2$ where
         $\omega_4(n)= \left\langle\left(\sum_{j}
             |W_{n,j}|^2\right)^2\right\rangle$
         of the shell vorticity distribution
         function. The horizontal line is the prediction if the
         three shell variables are assumed to be uncorrelated
         Gaussian variables with the same variance.
        }
\label{f:kurtosis}
\end{figure}

\begin{figure}
\caption{The mean enstrophy flux (triangle), and the mean squared
         enstrophy flux (square), in logarithmic scale. At the peak
         the standard deviation of the flux is about $50$ times larger
         than the mean value.
        }
\label{f:enstrophy_flux}
\end{figure}

\begin{figure}
\caption{The mean energy flux (triangle), and the mean squared
         energy flux (square), in logarithmic scale. At the peak
         the standard deviation of the flux is about $10,000$ times larger
         than the mean value.
        }
\label{f:energy_flux}
\end{figure}

\begin{figure}
\caption{Estimates of $\kappa_1$ for a system of $50$ shells from $-24$
         to $+25$.
         The symbols refer to runs with forcing (\protect\ref{fluxconst})
         $\nu = 10^{-9}$, $\nu' = 10^{-6}$ and flux = $0.05$ (square);
         $\nu = 10^{-9}$, $\nu' = 10^{-6}$ and flux = $0.1$ (triangle);
         $\nu = 10^{-6}$, $\nu' = 10^{-4}$ and flux = $0.2$ (full square).
         The full triangles refer to the data of figure
         \protect\ref{f:stateq}.
        }
\label{f:gaskinetics}
\end{figure}

\begin{figure}
\caption{Lyapunov exponent as a function of Reynolds number, defined
         as $Re = \nu^{-1}$.
         The system consists of $50$ shells from $-24$ to $+25$
         with forcing (\protect\ref{fluxconst}) with
         $\eta = 0.1$ and $\nu' = 10^{-6}$.
        }
\label{f:Lyapunov}
\end{figure}

\begin{figure}
\caption{Spreading of a perturbation as a function of time.
         The bottom curve refers to time $0.2$ seconds from
         the perturbation. The others, bottom to top,
         after time intervals of $0.4$ seconds.
         The system consists of $50$ shells from $-24$ to $+25$
         with forcing (\protect\ref{fluxconst}) with
         $\eta = 0.1$, $\nu = 10^{-9}$ and $\nu' = 10^{-6}$.
         At the initial time the perturbation was of strength
         $10^{-10}$.
         The curves are rather insensitive to the location of the initial
         perturbation, in figure it was located on shells $-6$ and $-5$.
        }
\label{f:spreading}
\end{figure}


\begin{references}

\bibitem{Obukhov71}
         A.M Obukhov,
         Atmos. Oceanic Phys. {\bf 7}, 41 (1971).

\bibitem{DesnyanskiNovikov}
         V.N Desnianskii and E.A. Novikov,
         J. of Appl. Mathem. and Mech., PMM {\bf 38}, 468 (1972).

\bibitem{Siggia77}
         E.D. Siggia,
         Phys. Rev. A {\bf 15}, 1730 (1977).

\bibitem{Zimin}
         V.D. Zimin,
	 Izv. AN SSSR, Fizika Atmosfery i Okeana {\bf 17}, 941 (1981).

\bibitem{YamadaOhkitani87}
         M. Yamada K. Ohkitani,
         J. Phys. Soc. of Japan {\bf 56}, 4210 (1987);
         M. Yamada and K. Ohkitani, Phys. Rev. Lett. {\bf 60}, 983 (1988).

\bibitem{JensenPaladin91}
         M.H. Jensen, G. Paladin and A. Vulpiani,
         Phys.Rev. A {\bf 43}, 798 (1991);
         M.H. Jensen, G. Paladin and A. Vulpiani,
         Phys.Rev. A {\bf 45}, 7214 (1992).

\bibitem{EggersGrossmann}
         J. Eggers and S. Grossmann,
         Physics Lett. A {\bf 156}, 444 (1991).

\bibitem{Gledzer}
         E.B. Gledzer,
         Sov.Phys.Dokl. {\bf 18}, 216 (1973).

\bibitem{Pisarenko93}
         D. Pisarenko, L. Biferale, D. Courvoisier, U. Frisch,
         M. Vergassola,
         Phys. Fluid A {\bf 5}, 2533 (1993).

\bibitem{Biferale94}
         L. Biferale, A. Lambert, R. Lima and G. Paladin,
         ``Transition to chaos in a shell model of
	 turbulence'', (preprint 1994).

\bibitem{Onsager}
         L. Onsager,
         Nuovo Cimento Suppl. {\bf 6}, 279 (1949).

\bibitem{Hopf}
         E. Hopf,
         J. Rat. Mech. Anal. {\bf 1}, 87 (1952).

\bibitem{Lee}
         T.D. Lee,
         Q. Appl. Math. {\bf 10}, 69 (1952).

\bibitem{KraichnanMontgomery}
         R.H. Kraichnan and D. Montgomery,
         Rep. Prog. Phys. {\bf 43}, 547 (1980).

\bibitem{YamadaOhkitani88a}
         M. Yamada and K. Ohkitani,
         Progr. Theo. Phys. {\bf 79}, 1265 (1988).

\bibitem{Frick}
         P.G. Frick,
         Magnetohydrodynamics {\bf 19:1},  48 (1983)
         [Translated from Russian].

\bibitem{FrickAurell}
         P.G. Frick and E. Aurell,
         Europhys. Lett. {\bf 24}, 725 (1993).

\bibitem{FrischSulem}
         U. Frish and P. Sulem,
         Phys. Fluids {\bf 27}, 1921 (1984).

\bibitem{Borue}
         V. Borue,
         Phys. Rev. Lett. {\bf 72}, 1475 (1994).

\bibitem{Kolmogorov41}
        A.N Kolmogorov,
        C.R. Acad. Sci. USSR {\bf 30}, 301 (1941).

\bibitem{Batchelor}
         G. Batchelor,
         Phys. of Fluids Suppl. II {\bf 12}, 233. (1969).

\bibitem{Kraichnan}
         R.H. Kraichnan,
         Phys. of Fluids {\bf 10}, 1417 (1967).

\bibitem{MoninYaglom}
         A. Monin and A. Yaglom,
         {\it Statistical Fluid Mechanics},
	 (MIT Press, 1971).

\bibitem{Anselmet84}
         F. Anselmet, Y. Gagne, E.J. Hopfinger and R. Antonia,
         J. Fluid Mech. {\bf 140}, 63 (1984).

\bibitem{Castaing90}
         B. Castaing,  Y. Gagne and E.J. Hopfinger,
         Physica D {\bf 46}, 177 (1990).

\bibitem{BasdevantCouder}
         C. Basdevant and Y. Couder,
         J. Fluid Mech. {\bf 173}, 225 (1986).

\bibitem{Benzi84}
         R. Benzi, G. Paladin, G. Parisi and A. Vulpiani,
         J. Phys. A {\bf 17}, 3521 (1984).

\bibitem{SmithYakhot}
         M. Smith and V. Yakhot,
	 Phys. Rev. Lett. {\bf 71}, 352 (1993).

\bibitem{OhkitaniYamada}
         K. Ohkitani and M. Yamada,
         Progr. Theo. Phys. {\bf 81}, 329 (1989).

\bibitem{LandauLifshitz10}
         L. Landau and E. Lifshitz,
         {\it Course on Teoretical Physics: Physical Kinetics},
         vol. 10, (Pergamon Press, 1981).

\bibitem{Boffettaunpub}
         G.  Boffetta,
         (unpublished, 1994).

\bibitem{Oseledec}
         V.I. Oseledec,
         Trans. Moscow Math. Soc. {\bf 19}, 197 (1968).

\bibitem{Benzi85}
         R. Benzi, G. Paladin, G. Parisi and A. Vulpiani,
         J. Phys. A {\bf 18}, 2157 (1985).

\bibitem{crisanti93}
         A. Crisanti, M.H. Jensen, G. Paladin and A. Vulpiani,
         Phys. Rev. Lett. {\bf 70}, 166 (1993);
         A. Crisanti, M.H. Jensen, G. Paladin and A. Vulpiani,
         J. Phys A {\bf 26}, 6943 (1993).

\bibitem{Ruelle}
         D. Ruelle,
         Comm. Math. Phys. {\bf 87}, 287 (1982).

\bibitem{Lorenz69}
         E. Lorenz,
         Tellus {\bf 21}, 3 (1967).

\bibitem{Benzi91}
         R. Benzi, L. Biferale, G. Paladin, A. Vulpiani
         and M. Vergassola,
         Phys. Rev. Lett. {\bf 67}, 2299 (1991).

\bibitem{FrischVergassola}
         U. Frish and M. Vergassola,
         Europhys. Lett. {\bf 15}, 439 (1991).

\bibitem{Aurell}
         E. Aurell, P. Frick and V. Shaidurov,
         Physica D {\bf 72}, 95 (1994).
\end{references}
\end{document}